\documentclass[]{aipproc}

\layoutstyle{6x9}

\SetInternalRegister\hbadness{8000} 

\begin{document}
\begin{flushright}
SLAC-PUB-9009\\
November 2001
\end{flushright}

\title{Testing Factorization 
\thanks{Invited Talk at the 9th International
Symposium on Heavy Flavor Physics, Caltech,
Pasadena, September 10-13, 2001.}
}


\author{Gudrun Hiller 
\thanks{Work supported by the Department of Energy, 
Contract DE-AC03-76SF00515} }
{address={Stanford Linear Accelerator Center, 
              Stanford University, Stanford, CA 94309, USA }}

\copyrightyear  {2001}

\begin{abstract}
We briefly review the status of factorization in $b$-decays.
We discuss several experimental tests of its nature and 
stress their importance.
We show that decays into mesons which have small 
decay constants or spin greater than one (`designer mesons') offer 
a variety of new opportunities.
\end{abstract}

\date{\today}

\maketitle

\section{Introduction}

Major theory issues in $b$-physics are
1) Is the CKM description of CP violation correct or are there other
sources of CP violation ?
and
2) Is the Standard Model the correct effective theory up scales of order
$\sim 1$ TeV ? The latter 
could be probed for example with Flavor-changing-neutral
current (FCNC) decays such as
$b \to s \gamma$, $b \to s \ell^+ \ell^-$ and $b \to s q \bar{q}$.
Both questions can be addressed in $b$-physics in a unique way, which has
stimulated many experimental and
theoretical activities.
However, there is another one
3) Is our understanding of non-perturbative QCD good enough to answer
the above questions ?
which is part of the whole picture.
Among effects due to hadronization, those related to the 
factorization of matrix
elements of hadronic 2-body decays are of peak
importance. Their understanding and knowledge of the limitations in their 
theoretical description
will become more urgent in the near future since
the `QCD background' limits the potential of precision tests of the 
Standard Model.
Fortunately, there are many decay modes, observables and
facilities where this can be further explored and checked.

\section{Factorization concept}

Naive factorization is a working hypothesis \cite{BSW}, which allows one
to express the matrix elements of hadronic 
2-body decays in terms of known objects, with the product
(decay constant $\times$ form factor).
Diagrammatically, it amounts to cutting the amplitude across the
$W$-boson line and resembles the description of semileptonic decays.
The picture of color transparency (CT) 
\cite{colortransy} 
justifies this for some decays
when the meson emitted from the weak vertex 
is fast in the rest frame of the decaying
parent. An example is $\bar{B}_0 \to D^+ \pi^-$.
The CT explanation however must fail e.g.~for
$\bar{B}_0 \to D^- \pi^+$.
If $1/N_c$ counting arguments \cite{1/N} are at work, 
then factorization would still hold here.
There exist no general theory for all 2-body decays, 
but they have been classified into
color allowed, suppressed, heavy-light, etc.
QCD based approaches \cite{BBNS,pQCD} differ in their treatment
of $\alpha_s$ and $1/m_b$ power corrections.
Thus it is important to test experimentally where 
(naive) factorization holds and where
corrections arise to understand its dynamical origin
(CT, $1/N_c$, ... ).

\subsection{status}

Factorization has been tested
with tree level dominated modes, i.e.~where possible New Physics effects
are tiny.
Currently, the factorization concept in color allowed $B$-decays rules:
\begin{itemize}
\item Heavy-to-light $\bar{B}_0 \to D^{(*)+} (\pi, \rho,a_1)^-$
decays can be described by one universal coefficient
$|a_1|=1.1 \pm 0.1$, see e.g.~\cite{BBNS}; factorization is ok up to
the ${\cal{O}}(10 \%) $ level.
\item In $\bar{B}_0 \to D^{(*)+} X^-$ decays, where $X=4 \pi, \omega
\pi$ the factorization hypothesis can be tested as a function of
the hadronic mass $m_X$ of the emitted $n \pi$ state \cite{LLW01}. Its 
`decay constant' $\langle X| \bar{d} \Gamma u|0 \rangle$ 
is obtained from hadronic $\tau$-data.
CT holds if $X$ is fast, so one 
expects corrections to factorization for growing $m_X$, but no
kinematical dependence in the $1/N_c$ approach.
Current data indicate no factorization breaking but further
experimental studies should be pursued.
\item In $\bar{B}_0 \to D_{(s)}^- \pi^+$ decays there is no CT 
explanation of factorization.
The factorizable amplitude is proportional to
${\cal{A}}_{fac} \sim V_{ub} F^{B \to \pi}(m_{D(s)}^2) f_{D_s}$ and
quantitative tests need good control over all parameters on the r.h.s.
Experimentally, to date an upper bound exists,
${\cal{B}}(\bar{B}_0 \to D_s^- \pi^+) < 1.1 \cdot 10^{-4}$ $@ 90\%$ 
C.L.~\cite{belleconf2001} which is ok with factorization.
\item In heavy-to-heavy $B \to D^{(*)} D_s^{(*)}$ decays 
factorization holds within errors \cite{luorosner}; confirmed by
study including penguins  \cite{Kim:2001cj}. The
dominant uncertainty in these analyses comes from the decay constants,
e.g.~$f_{D_s}=280 \pm 48$ MeV from $D_s \to \mu \nu_{\mu}$
\cite{Groom:2000in}.
\end{itemize}
Significant improvement in precision is required to isolate
factorization breaking effects in the above decays.
Alternatively, one can study factorization
in those decays, where the corrections to factorization are not hidden
behind a large factorizable contribution \cite{explorefac,babarbook}.
Then less precision is required, although for the price of less events.

\section{New Ways}

\begin{table}
\begin{tabular}{cccccccccccc}
\hline
    \tablehead{1}{c}{b}{$X$}
  & \tablehead{1}{c}{b}{$a_0$}
  & \tablehead{1}{c}{b}{$b_1$}
  & \tablehead{1}{c}{b}{$\pi$}
  & \tablehead{1}{c}{b}{$a_2$}
  & \tablehead{1}{c}{b}{$a_0$}
  & \tablehead{1}{c}{b}{$\pi_2$}
  & \tablehead{1}{c}{b}{$\rho_3$}
  & \tablehead{1}{c}{b}{$\chi_{c0}$}
  & \tablehead{1}{c}{b}{$K_0^*$} 
  & \tablehead{1}{c}{b}{$K_2^* $} 
  & \tablehead{1}{c}{b}{$D_2^* $} \\
\hline
$m_X$ [MeV]  & 985     & 1230     & 1300     & 1318     & 1474     &
              1670     & 1691      &   3415   & 1412  & 1426 & 2459   \\
$J^{PC}$ &  $0^{++}$ & $1^{+-}$ & $0^{-+}$ & $2^{++}$ & $0^{++}$ & 
              $2^{-+}$ & $3^{--}$ & $0^{++}$ &$0^{+}$& $2^{+}$ &
              $2^{+}$  \\
\hline
\end{tabular}
\caption{ Examples of `designer' mesons.}
\label{tab:friends}
\end{table}

\begin{figure}
 \makebox{\includegraphics[height=.2\textheight]{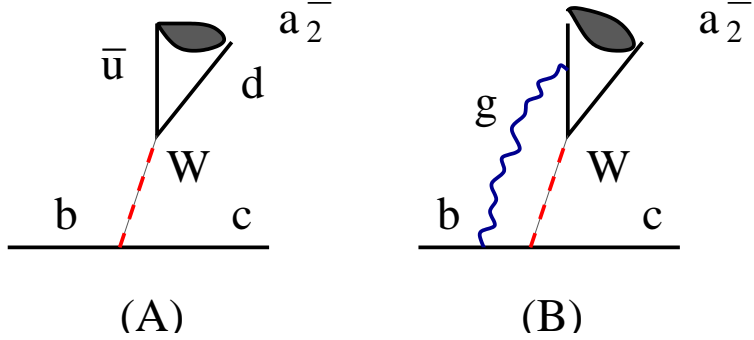}}
  \caption{Examples of a diagram which does not (A) and does
(B) contribute to $b \to c a_2^-$.}
\label{fig:tree}
\end{figure}

Recently, it has been proposed to explore factorization with
final states whose coupling to the $W$-boson is suppressed
either because the spin $>1$ or the decay constant is
suppressed \cite{explorefac}.
Examples of such `designer mesons' are given in Table \ref{tab:friends}.
The relevant property in the context of factorization tests is that
amplitudes in {\it naive} factorization are suppressed. In the
case of a vanishing decay constant, they even vanish
e.g. ${\cal{B}}_{naive}(\bar{B}_0 \to D^+ a_2^-) =0$, 
see Fig.~\ref{fig:tree} (A).
However, corrections e.g.~induced by hard gluon exchange,
see Fig.~\ref{fig:tree} (B) circumvent 
suppression, because the $a_2^-$ is now produced
from a non-local vertex, which allows for higher spins.
Other mechanisms that avoid the suppression include annihilation 
topologies, interactions with the spectator and those induced by
charm \cite{charmstuff}.
Thus non-factorizable effects can be highlighted by the choice of specific
`designer' final states.

\subsection{decay constants}

Decay constants for scalars $S$ with momentum $p$ are defined as
$\langle S(p) |\bar{q} \gamma_\mu q^\prime |0 \rangle=-i f_S p^\mu$.
For $q=q^\prime$ the decay constant vanishes 
by charge conjugation e.g.~$f_{a_0^0}, f_{\chi_{c0}}=0$. The decay constant
of the charged $a_0$
is proportional to $m_d-m_u$ and thus
isospin suppressed and small compared to e.g.~$f_\pi =131$ MeV.
Analogous arguments apply to the axial vector $b_1$.
Also a $1^+$ single charm meson $D_{J=1}(j=3/2)$ with
vanishing decay constant in the heavy quark limit is predicted \cite{heavy}.
The decay constants of the mesons in Table \ref{tab:friends} are 
only poorly known. 
The current theory spread as compiled in \cite{explorefac} reads as
\begin{eqnarray}
f_{a_0(980)}   &=& 0.7 -2.5 \, \mbox{MeV} \nonumber \\
f_{\pi(1300)}  &=& 0.5 -7.2 \, \mbox{MeV} \nonumber \\
f_{K_0^*}      &=& 33 -46  \; \; \;  \mbox{MeV} 
\end{eqnarray}
As expected, $f_{K_0^*} > f_{a_0}$ due to larger quark mass splitting.
No estimate of the $b_1$ (longitudinal) 
decay constant has been reported. 
The decay constants of $a_0,\pi(1300)$, $b_1,K_0^*$ mesons could be
determined in hadronic $\tau$-decays. 
Estimates of the corresponding 
branching ratios are given in Table \ref{tab:tau}. Note that the bound 
${\cal{B}}(\tau \to (\pi(1300)\to 3 \pi) \nu_\tau) < 1 \cdot 10^{-4}$ 
\cite{Groom:2000in} implies $f_{\pi(1300)}<8.4$ MeV.

\begin{table}
\begin{tabular}{ccccc}
\hline
    \tablehead{1}{c}{b}{$X$}
  & \tablehead{1}{c}{b}{$a_0(980)$}
  & \tablehead{1}{c}{b}{$a_0(1450)$}
  & \tablehead{1}{c}{b}{$\pi(1300)$}
  & \tablehead{1}{c}{b}{$K_0^*(1430)$}   \\
\hline
$f_X$ [MeV] & 1.1 & 0.7 & 7.2   & 42 \\
${\cal{B}} (\tau \to \nu_\tau X)$ & $3.8 \cdot 10^{-6}$ & $3.7 \cdot 10^{-7}$ 
& $7.3 \cdot 10^{-5}$   & $7.7 \cdot 10^{-5}$ \\
\hline
\end{tabular}
\caption{Theory estimates of decay constants as complied in
\cite{explorefac} and the corresponding $\tau \to \nu_\tau X$ 
branching ratios.}
\label{tab:tau}
\end{table}

\subsection{flavor selection}

Not every decay into a designer final state is suppressed. 
Instead one needs flavor selection criteria to find the designer modes.
It is crucial that the spectator does not end up in the designer.
An example where this condition does not hold is the decay 
$B^- \to D^0 a_0^-$. It proceeds via two different 
topologies. The color allowed one has a suppressed amplitude, because
the $a_0^-$ is emitted from the weak vertex. The color suppressed
contribution to the amplitude however produces the $a_0^-$
from the spectator. Since form factors for designer mesons are not 
anomalous, this topology escapes suppression.
Examples of modes which satisfy the criteria and do have a suppressed
factorizable contribution are given in Table~\ref{tab:flavor},
(for the full listing and details see \cite{explorefac}).
There are  
many decays of $B_0,B^\pm,B_s$ mesons and $\Lambda_b,\Omega_b$ baryons,
many final states, which cover a wide range of masses
(every particle can be replaced by another one with the same
flavor content and $W$-coupling features),
many topologies (tree, annihilation, penguin annihilation,...), and
many classifications (color allowed, color suppressed).
Heavy-light, light-light and decays into charmonium should factorize 
according to CT while light-heavy and heavy-heavy not.
Note that baryons differ in quark content and in particular in
annihilation topologies from mesons, i.e.~they cannot be fully
annihilated by 4-Fermi operators. They also offer more degrees of
freedom accessible to experiments such as polarization and have no 
background from the decay of the CP conjugate parent into the same
final state (this is a problem for e.g.~$\bar{B}_0 \to \pi^+ a_0^-$,
because $B_0 \to \pi^+ a_0^-$ is unsuppressed).

\begin{table}
\begin{tabular}{lccccc} 
 \hline 
    \tablehead{1}{l}{t}{example decay}
 &  \tablehead{3}{c}{t}{factorizing contribution\\
quark level \hspace{0.5cm} tree \hspace{0.5cm} penguin}
 &  \tablehead{2}{c}{t}
   {annihilation\\ tree \hspace{0.6cm} penguin } \\
\hline
$\bar{B}^0 \to D^{+} a_0^-$   & 
  $\bar{d}b\to \bar{d}(c\bar{u}d)$ &
  $\lambda^2$ &  & $\lambda^2$ & \\
$B^- \to \pi^0 D_2^{*-}$ &        
  $\bar{u}b\to \bar{u}(u\bar{c}d)$ & 
  $\lambda^4$ & & $\lambda^4$ & \\
$\bar{B}_s \to D_s^{+} D_2^{*-}$ &      
  $\bar{s}b\to \bar{s}(c\bar{c}d)$ & 
  $\lambda^3$ & $\lambda^3$ &  & $\lambda^3$ \\ 
$B^-\to \pi^- \bar{K}_2^{*0}$ & 
  $\bar{u}b\to \bar{u}(d\bar{d}s)$ &
   & $\lambda^2$ &  $\lambda^4$ & $\lambda^2$ \\
$B^-\to K^- K_2^{*0}$ & 
  $\bar{u}b\to \bar{u}(s\bar{s}d)$ &
   & $\lambda^3$  & $\lambda^3$ & $\lambda^3$ \\
\hline 
$\bar{B}^0 \to \pi^0 D_2^{*0}$ & 
  $\bar{d}b \to \bar{d}(d\bar{u}c)$ & 
  $\lambda^2$ &  & $\lambda^2$ & \\
$B^-\to K^- \chi_{c0}$ & 
  $\bar{u}b\to \bar{u}(s\bar{c}c)$ &
  $\lambda^2$ & $\lambda^2$ & $\lambda^4$ & $\lambda^2$ \\
$\bar{B}_s \to K^0\, a_2^0$ & 
  $\bar{s}b\to \bar{s}(d\bar{u}u)$ &
  $\lambda^3$ & $\lambda^3$ & & $\lambda^3$ \\
\hline
$\Lambda_b\to \phantom{\Lambda_c K_0^{*-},\;}
  n D_2^{*0}$ &
  $udb\to ud(c\bar{u}d)$ & 
  $\lambda^2$ & & $\lambda^2$\\
$\Lambda_b\to \Lambda_c D_{sJ}^-,\, \Lambda \chi_{c0}$ &
  $udb\to ud(c\bar{c}s)$ & 
  $\lambda^2$ & $\lambda^2$ & 
  $\lambda^4$ & $\lambda^2$  \\
$\Omega_b\to \Omega_c a_0^{-},\, \Xi^- D_2^{*0}$ &
  $ssb\to ss(c\bar{u}d)$ &
  $\lambda^2$ & \\
\hline
\end{tabular}
\caption{\label{tab:flavor} Some 
color allowed, color suppressed and baryon decays which satisfy the
flavor selection criteria specified in text, adapted 
from \cite{explorefac}. The
magnitude of the amplitudes is given in powers of the 
Wolfenstein parameter $\lambda\simeq 0.22$.}
\end{table}

\subsection{existing data}

Experimental information on some designer modes from
Table \ref{tab:flavor} is already available.
The Belle collaboration has reported recently \cite{Bellechi0}
\begin{eqnarray}
{\cal{B}}(B^+ \to \chi_{c0} K^+)&=&(6.0^{+2.1}_{-1.8} \pm 1.1) \cdot
10^{-4} \nonumber \\
R=\frac{{\cal{B}}(B^+ \to \chi_{c0} K^+)}
       {{\cal{B}}(B^+ \to J/\Psi K^+)}&=&0.60^{+0.21}_{-0.18} \pm 0.05 \pm 0.08
\label{eq:violation}
\end{eqnarray}
Because it has a small radius, CT is expected to work for charmonium
despite the fact it is not light. 
However, problems of factorization 
with color suppressed decays are no surprise, since
radiative corrections come in without color suppression and are
large \cite{BBNS,explorefac}.
Indeed, eq.~(\ref{eq:violation}) represents an ${\cal}{O}(1)$ 
violation of naive factorization since $R_{naive}=0$.

The branching ratio into a light-light final state has 
been measured by Belle 
${\cal{B}}(B^+ \to (K_X(1400) \to K^+ \pi^-) \pi^+)=
(12.7^{+3.5+1.8+2.9}_{-3.4-1.8-5.8}) \cdot 10^{-6}$ \cite{Abe:2001pg}.
This is comparable in magnitude with the one into the
corresponding unsuppressed mode, 
${\cal{B}}(B^+ \to K^0  \pi^+)=
(13.7^{+5.7+1.9}_{-4.8-1.8}) \cdot 10^{-6}$ (Belle) \cite{BelleKpi} and 
${\cal{B}}(B^+ \to K^0  \pi^+)=
(18.2^{+3.3}_{-3.0} \pm 2.0) \cdot 10^{-6}$ (BaBar) \cite{Aubert:2001hs}.
Both decays are dominated by QCD penguins. In particular, they receive
large contributions from scalar penguins
$(\bar{q} (1-\gamma_5) b)(\bar{s} (1+\gamma_5) q)$, 
which are parametrically enhanced by factors
\begin{eqnarray}
\nonumber
r_K=\frac{2 m_K^2}{m_b m_s}\; 
,~~~~~~~~~~ r_{K_0^*}=\frac{2 m_{K_0^*}^2}{m_b m_s}
\end{eqnarray}
for pseudoscalar $K$ and scalar $K_0^*$ mesons, respectively.
Since the penguin enhancement $r_{K_0^*}/r_K =m_{K_0^*}^2/m_K^2 $ 
compensates for the decay constant suppression
$f_{K_0^*}/f_K \sim 1/4$,
the hypothesis $K_X(1400) = K_0^*$ is consistent with factorization
\cite{explorefac,chernyak}.
Note that both contributions remain finite
in the chiral limit -- for the Goldstone bosons because $m_K \to 0$ in
the same limit and for the scalars because 
$f_{K_0^*} \sim m_s$ which multiplies $r_{K_0^*}$ vanishes.
Measurement of the branching ratio into the tensor
would be much more exciting since in naive factorization
${\cal{B}}_{naive}(B^\pm \to K_2^* \pi^\pm)=0$.
Thus, one would directly probe the
factorization breaking corrections, an issue in light-light decays 
that is controversial between perturbative QCD \cite{pQCD}
and QCD factorization \cite{BBNS,BBNS2} 
(Some problems in the pQCD approach have been 
pointed out recently in Ref.~\cite{problems}).
Experimentally, angular analysis is required to discriminate the nearby kaon
resonances $K_0^*,K^*(1410),K_2^*$.

\subsection{quantitatively: color allowed $B \to D^{(*)} X$ decays\\
$X=a_0,a_2,b_1,\pi(1300),\pi_2,\rho_3$ and $K_0^*,K_2^*$}

Generically we have the branching ratios 
${\cal{B}}(B \to D^{(*)} (\pi,\rho,a_1)) \sim 10^{-3}$.
Assuming ${\cal{O}}(10) \%$ corrections to factorization 
arising from $1/N_c^2 $ and/or $\Lambda_{QCD}/m_b$ 
one expects for the $I=1$ designer mesons 
${\cal{B}}(B \to D^{(*)} X) \sim 10^{-5}-10^{-6}$.
The branching ratios can be calculated in QCD factorization
\cite{BBNS} from evaluation of the matrix element \cite{explorefac}
\begin{eqnarray}
\langle D^{(*)+} X^-| {\cal{H}}_{eff}|B \rangle \sim
a_1(\mu) f_X + \frac{\alpha_s(\mu)}{4 \pi} C_2(\mu) \frac{C_F}{N_c}
\int_0^1 du F(u) \varphi(u;\mu)
\label{eq:qcd}
\end{eqnarray}
The first term corresponds to the expression in naive factorization.
It vanishes if $f_X=0$.
In terms of light cone distribution amplitudes (DA) $\varphi(u;\mu)$, 
where $u$ denotes the momentum fraction carried by the quark in $X$,
the decay constant is given as
\begin{eqnarray}
\nonumber
f_X =\int_0^1 du \varphi(u)
\end{eqnarray}
However, a small  or vanishing zeroth moment $f_X \simeq 0$ does not imply
that the DA is small or vanishing.
The contribution from hard gluon exchange, which is given by the
second term in eq.~(\ref{eq:qcd}), thus escapes the suppression
mechanism \cite{explorefac}.
{}From charge conjugation, the following symmetry properties for meson
DA's hold up to isospin breaking
\begin{eqnarray}
\pi,\pi(1300),\pi_2,\rho_3 &:& \varphi(u)=+\varphi(1-u) \nonumber \\
a_0,a_2,b_1,K_0^*,K_2^* &:&\varphi(u)=-\varphi(1-u)
\end{eqnarray}
The leading coefficients in the expansion of the DA in Gegenbauer 
polynomials $C_n^{3/2}$
\begin{eqnarray}
\varphi(u;\mu) = f^\varphi\, 6 u(1-u) \Big[ B_0 + \sum_{n=1}^{\infty}
              B_n(\mu)\, C_n^{3/2}(2u-1) \Big] 
\nonumber
\end{eqnarray}
are estimated for mesons $X$ 
with Fock space normalization techniques a la pion as 
\begin{eqnarray}
|f^\varphi B_1|_{a_0, b_1, a_2, K_0^*, K_2^*} 
      &\approx& 75~\mbox{MeV} 
\nonumber \\
|f^\varphi B_2|_{\pi(1300), \pi_2, \rho_3} &\approx& 50~\mbox{MeV} 
\end{eqnarray}
at the renormalization scale $\mu=m_b=4.4$~GeV 
($B_0=1$ and $f^\varphi=f_X$ for spin$(X) \leq1$, otherwise $B_0=0$)
\cite{explorefac}.
Resulting branching ratios are given in Table \ref{tab:br}.
\begin{table}
\begin{tabular}{lccc}
\hline
    \tablehead{1}{l}{t}{mode}
 & \tablehead{1}{c}{t}{naive factorization}
 & \tablehead{2}{c}{t}
   {QCD factorization\\$\mu=m_b$ \hspace{1cm} $\mu=m_b/2$ } \\
\hline
$\bar{B}^0 \to D^+ a_0(980)$ & $1.1\cdot 10^{-6}$
                             & $2.0\cdot 10^{-6}$ & $4.0\cdot 10^{-6}$
\\
$\bar{B}^0 \to D^+ a_0(1450)$ & $8.6\cdot 10^{-8}$
                              & $5.8\cdot 10^{-7}$ & $2.1\cdot 10^{-6}$
\\
$\bar{B}^0 \to D^+ a_2,\, D^+ b_1$$^\dagger$ & 0
                              & $3.5 \cdot 10^{-7}$ & $1.7 \cdot 10^{-6}$
\\
$\bar{B}^0 \to D^+ \pi(1300)$ & $9.1\cdot 10^{-6}$
                              & $9.3\cdot 10^{-6}$ & $9.6\cdot 10^{-6}$
\\
$\bar{B}^0 \to D^+\pi_2,\, D^+\rho_3$ & 0
                              & $1.4 \cdot 10^{-9}$ & $8.1 \cdot 10^{-9}$
\\
$\bar{B}^0 \to D^+ K^*_0(1430)$ & $2.0\cdot 10^{-5}$ & $2.0\cdot 10^{-5}$
                                & $2.1\cdot 10^{-5}$
\\
$\bar{B}^0 \to D^+ K^*_2$ & 0 & $1.9\cdot 10^{-8}$ & $9.2\cdot
    10^{-8}$ \\
\hline
$\bar{B}^0 \to D^{*+} a_0(980)$ & $1.0\cdot 10^{-6}$
                                & $1.8\cdot 10^{-6}$ & $3.7\cdot 10^{-6}$
\\
$\bar{B}^0 \to D^{*+} a_0(1450)$ & $7.9\cdot 10^{-8}$
                                 & $5.2\cdot 10^{-7}$ & $1.9\cdot 10^{-6}$
\\
$\bar{B}^0 \to D^{*+} a_2 \, D^{*+} b_1$$^\dagger$ & 0
                                 & $2.9 \cdot 10^{-7}$ & $1.5 \cdot 10^{-6}$
\\
$\bar{B}^0 \to D^{*+} \pi(1300)$ & $8.3\cdot 10^{-6}$
                                 & $8.4\cdot 10^{-6}$ & $8.4\cdot 10^{-6}$
\\
$\bar{B}^0 \to D^{*+}\pi_2,\, D^{*+}\rho_3$ & 0
                                 & \, $5.7\cdot 10^{-10}$ & $3.2\cdot 10^{-9}$
\\
$\bar{B}^0 \to D^{*+} K^*_0(1430)$ & $1.8\cdot 10^{-5}$
                                   & $1.9\cdot 10^{-5}$ & $1.9\cdot 10^{-5}$
\\
$\bar{B}^0 \to D^{*+} K^*_2$ & 0 & $1.5\cdot 10^{-8}$ & $7.7\cdot 10^{-8}$
\\ \hline
\end{tabular}
\caption{Branching ratios 
obtained in QCD factorization for two choices of the renormalization
scale $\mu$ and in the naive factorization approach, taken from
\cite{explorefac}. $^\dagger$With $f_{b_1}=0$.}
\label{tab:br}
\end{table}
Similar ones are expected for $B_s \to D_s X$ decays 
accessible at the Tevatron and LHC.
The QCD factorization result shows an enhancement over the one obtained
in the naive factorization approach. However, the branching ratios
are still much smaller than the ones of the corresponding
`non-designer' modes like $B \to D \pi$, but within
experimental reach.
Since QCD factorization picks up hard $\alpha_s$ contributions 
which are enhanced in the designer decays, their 
amplitude is more sensitive to the renormalization scale and gains large strong
phases.
Current dominant uncertainties are due to the decay constants, 
which could be measured in hadronic $\tau$
decays, see Table \ref{tab:tau}.  Furthermore, information on 
the DA's of the neutral $a_0,a_2, \pi(1300),\pi_2$ could be
obtained from $\gamma \gamma^*$ collisions in
$e^+ e^- \to e^+ e^- X$ processes similar to the analysis performed
by CLEO for $\pi, \eta, \eta^\prime$ \cite{CLEOpi}.

\subsection{measure strong phases as a byproduct of $2 \beta + \gamma $}

Time dependent measurements in $B \to D^\pm \pi^\mp$ decays
allow the extraction of CKM angles $2 \beta + \gamma $
together with strong phases
without model assumptions on strong dynamics \cite{isi}.
Unfortunalty, the effect scales with a tiny $\sim 1\%$ asymmetry.
Choosing instead $B \to D^\pm X^\mp$ decays, where
$X=a_0,a_2,b_1,\pi(1300),\pi_2,\rho_3$ 
the CKM hierarchy of the amplitudes is compensated by 
the decay constants and large \cite{yet}, for example for the $a_0$
\begin{eqnarray}
\frac{{\cal{A}}( B_0 \to D^+ a_0^-)}{{\cal{A}}(\bar{B}_0 \to D^+
a_0^-)} \simeq 
\frac{f_D}{f_{a_0}} \frac{V_{cd} V_{ub}^*}{V_{cb} V_{ud}^*} \sim O(1)
\end{eqnarray}
Therefor designer mesons are
competitive with $B \to D^\pm \pi^\mp $ decays and
help to resolve ambiguities because different and large strong phases
are involved.

\section{Conclusions}

Factorization successfully works in color allowed $b$-decays
within present errors.
This includes $B \to D D_s$ decays where the color transparency
explanation is absent.
More data are needed to see to what accuracy it works and perhaps to  
understand why it works.
I discussed new strategies to explore 
factorization with modes where the factorizable contribution is
suppressed such that the corrections show up isolated.
This is similar to pure annihilation decays
$B_0 \to K^+ K^-$, $B_s \to \pi^+ \pi^-, \pi^0 \pi^0$ and
$B \to baryon \; baryon$.
These measurements could be carried out at
operating and future high luminosity $e^+ e^-$-facilities \cite{superbb}.
A dedicated factorization study
requires improved knowledge of non-perturbative input
such as decay constants and distribution amplitudes. 
Their determination should be part of such an experimental program.
A qualitatively and quantitatively accurate description of hadronic
matrix elements in $B$-decays 
is particularly important for the extraction of the CKM unitarity angles
$\gamma$ and $\alpha$.
Finally, `designer' final states and modes are also promising to study
CP violation and to search for New Physics effects in rare FCNC processes.
Steps in this direction have already been undertaken
\cite{yet,sandrine,babar_a0}.

\begin{theacknowledgments}
It is a pleasure to thank the organizers for the invitation to this 
exciting symposium and Dr.~Diehl for joyful collaboration on 
parts of the work presented here. 
I am grateful to David E.~Kaplan for useful comments on the manuscript.
\end{theacknowledgments}

\end{document}